\documentclass[%
 reprint,
superscriptaddress,
 amsmath,amssymb,
 aps,
]{revtex4-2}

\usepackage{graphicx}
\usepackage{dcolumn}
\usepackage{bm}
\usepackage{hyperref}

\newcommand{\myAuthor}[1]{#1}
\usepackage[dvipsnames]{xcolor}

\usepackage{acronym}
\usepackage{placeins}

\begin{document}


\title{
Self-organized defect clustering and concentration-dependent vacancy diffusion in MoS\textsubscript{2}
}

\author{Aaron Flötotto}
\email{aaron.floetotto@tu-ilmenau.de}
\affiliation{%
  Technische Universität Ilmenau, Institute of Physics,\\
  Weimarer Straße 25, 98693 Ilmenau, Germany
}%
\affiliation{
  Institute of Micro- and Nanotechnologies,\\
  Ehrenbergstraße 29, 98693 Ilmenau, Germany
}

\author{Benjamin Spetzler}
\affiliation{
  Christian-Albrechts-Universität zu Kiel, Institute of Materials Science,\\
  Kaiserstraße 2, 24143 Kiel, Germany
}%

\author{Martin Ziegler}
\affiliation{
  Christian-Albrechts-Universität zu Kiel, Institute of Materials Science,\\
  Kaiserstraße 2, 24143 Kiel, Germany
}%

\author{Erich Runge}
\affiliation{%
  Technische Universität Ilmenau, Institute of Physics,\\
  Weimarer Straße 25, 98693 Ilmenau, Germany
}%
\affiliation{
  Institute of Micro- and Nanotechnologies,\\
  Ehrenbergstraße 29, 98693 Ilmenau, Germany
}

\author{Christian Dreßler}
\email{christian.dressler@tu-ilmenau.de}
\affiliation{%
  Technische Universität Ilmenau, Institute of Physics,\\
  Weimarer Straße 25, 98693 Ilmenau, Germany
}%
\affiliation{
  Institute of Micro- and Nanotechnologies,\\
  Ehrenbergstraße 29, 98693 Ilmenau, Germany
}

\date{\today}

\begin{abstract}
    Sulfur vacancy migration has a crucial impact on electronic transport and the functional behavior of MoS\textsubscript{2}-based devices such as memristors and memtransistors.
    According to recent atomistic simulations, vacancy migration proceeds via cooperative, vacancy-assisted sulfur jumps, implying strongly correlated defect dynamics.
    Here, we investigate the collective behavior of sulfur-vacancy clusters in MoS\textsubscript{2} using \ac{KMC} simulations with transition rates derived from \ac{MLIP} \ac{MD} simulations.
    We identify three transport regimes: At low concentrations, vacancies are immobile or confined within small clusters, whereas at high concentrations, classical diffusive transport with a constant diffusion coefficient is observed, and vacancies aggregate into anisotropically extended clusters.
    A well defined intermediate regime is characterized by clusters merging into a connected, fluctuating network with a concentration-dependent diffusion coefficient.
    This regime is characterized by a broad distribution of cluster sizes.
    The strong dependence of the vacancy diffusion coefficient on the average defect concentration provides new insights into the origin of memristive behavior observed in MoS\textsubscript{2}.
\end{abstract}


\maketitle

\acrodef{MD}{molecular dynamics}
\acrodef{DFT}{density functional theory}
\acrodef{NEB}{nudged elastic band}
\acrodef{MLIP}{machine learning interatomic potential}
\acrodefplural{MLIP}[MLIPs]{machine learning interatomic potentials}
\acrodef{RDF}{radial distribution function}
\acrodef{MSD}{mean squared displacement}
\acrodefplural{MSD}[MSDs]{mean squared displacements}
\acrodef{AIMD}{\textit{ab initio} molecular dynamics}
\acrodef{GAP}{Gaussian approximation potential}
\acrodefplural{GAP}[GAPs]{Gaussian approximation potentials}
\acrodef{GPR}{Gaussian process regression}
\acrodef{ACE}{atomic cluster expansion}
\acrodef{GNN}{graph neural network}
\acrodef{SCF}{self-consistent field}
\acrodef{KMC}{kinetic Monte-Carlo}
\acrodef{MC}{Monte-Carlo}
\acrodef{TEM}{transmission electron microscopy}
\acrodef{FPT}{first-passage time}
\acrodefplural{FPT}[FPTs]{first-passage times}
\acrodef{PDF}{probability density function}
\acrodef{2D}{two-dimensional}
\acrodef{TMDC}{transition metal dichalcogenide}

\acresetall 
\section{Introduction}
\Ac{2D} molybdenum disulfide (MoS\textsubscript{2}) is a prototypical member of the \ac{TMDC} family and, arguably, the most extensively studied \ac{TMDC} for novel electronic applications\cite{Mak2010,Radisavljevic2011,Wang2012,Chhowalla2013}.    
The existence and migration of sulfur vacancies strongly affects material properties of MoS\textsubscript{2}, including catalysis, chemical sensing, photoluminescence, and electronic transport~\cite{Li2019,Hu2021,Fei-2023ADFM,Jin-2024ADFM,Chen2024,Pang2024}, and is therefore a key consideration in the design of MoS\textsubscript{2}-based devices.
Among a variety of electronic devices, memtransistors based on \ac{2D} MoS\textsubscript{2} are gaining increasing attention as next-generation computing devices~\cite{Sangwan2015,Li2018,Sangwan2018,Jadwiszczak2019,Wang2019,Lee2020,Feng2021,Yuan2021,Fu2023,MigliatoMarega2023,Zhu2023,Liu2024,Yang2024,Zou2024}.
Sulfur vacancy migration and the resulting fluctuation in the local vacancy concentration within the conduction channel are widely recognized as key factors that control the material’s intrinsic transport properties and the operational behavior of such devices~\cite{Sangwan2015,Li2018,Jadwiszczak2019,Li2021,Spetzler2022,Abdel2025,Spetzler2024,Spetzler2025,Wu2026}.

While the dynamics of individual sulfur vacancies~\cite{Komsa2013, Gao2021, Wang2022} and isolated vacancy clusters~\cite{Hlozny2025,Hlozny2026,Fltotto2026} have previously been studied using \ac{DFT} calculations and \ac{MLIP}-based \ac{MD} simulations, the collective dynamics of multiple vacancy clusters in MoS\textsubscript{2} and their dependence on vacancy concentration remain an important open question~\cite{Gali2020,Speckmann2025}. 
Addressing this problem in simulations at large time and length scales requires the inclusion of the recently identified cooperative vacancy migration mechanisms~\cite{Hlozny2025,Fltotto2026} into \ac{KMC} models.
Earlier \ac{KMC} studies~\cite{Wang2022, Wang2024} do not explicitly incorporate the cooperative dynamics within sulfur-vacancy clusters observed in the \ac{MD} simulations by us and others 
and, therefore, did not capture the concentration-dependent diffusion observed in the present work.

The sulfur-vacancy dynamics in MoS\textsubscript{2} are governed by sulfur-atom jumps ``assisted'' by a second, adjacent vacancy~\cite{Komsa2013, Gao2021, Wang2022, Hlozny2025, Hlozny2026, Fltotto2026}.
In the absence of such an assisting vacancy, sulfur migration into an isolated vacancy is suppressed by an energy barrier exceeding 2~eV~\cite{Komsa2013, Gao2021, Wang2022, Hlozny2026}.
When a second vacancy is present, the minimum-energy path passes through an interstitial site centered between three now-vacant sulfur lattice sites.
This cooperative jump mechanism lowers the energy barrier to 0.6--0.8~eV, allowing it to be thermally overcome at experimentally relevant temperatures, as shown in \ac{DFT} and \ac{MLIP} calculations~\cite{Komsa2013, Gao2021, Wang2022, Hlozny2025, Fltotto2026, Hanseroth2026, Hanseroth2026a} in agreement with \ac{TEM} measurements~\cite{Komsa2013,Zhou2017,Chen2018}.

The vacancy-assisted jump mechanism in MoS\textsubscript{2} bears similarities to mechanisms described in so-called kinetically constrained models, which have been put forward in the 1990's as models for glassy systems and remain relevant today~\cite{Ritort2003, Causer2020}. 
In particular, the kinetically constrained model introduced by \myAuthor{Jäckle} and \myAuthor{Kronig}~\cite{Jckle1994,Kronig1994} is qualitatively similar to the actual hopping mechanism observed in MoS\textsubscript{2}. 
They observe concentration-dependent transport and present arguments for a non-zero diffusion coefficient at all concentrations in the thermodynamic limit of an infinite lattice. 
Both findings would have implications for MoS\textsubscript{2}-based devices if they were transferable from this minimal kinetically constrained model to the actual cooperative vacancy dynamics in MoS\textsubscript{2}. 
However, while \myAuthor{Jäckle} and \myAuthor{Kronig} implemented a vacancy-assisted hopping rule on a triangular lattice that prohibits jumps of isolated vacancies, their model does not include MoS\textsubscript{2}-specific features that favor the formation of large vacancy clusters over their dissociation~\cite{Chen2018,Fltotto2026}. 
In addition, computational limitations at the time restricted the time and length scales accessible in Monte-Carlo simulations.

In the following, we derive \ac{KMC} rates from \ac{MLIP}-based \ac{MD} simulations, perform \ac{KMC} simulations of MoS\textsubscript{2} with initially random vacancy distributions to investigate self-organized clustering, and analyze the resulting concentration-dependent vacancy-diffusion coefficients.

\section{Results and Discussion}
\subsection{Transition rates from \ac{MLIP} \ac{MD} simulations}
\begin{figure}[htbp]
    \centering
    \includegraphics[width=\linewidth]{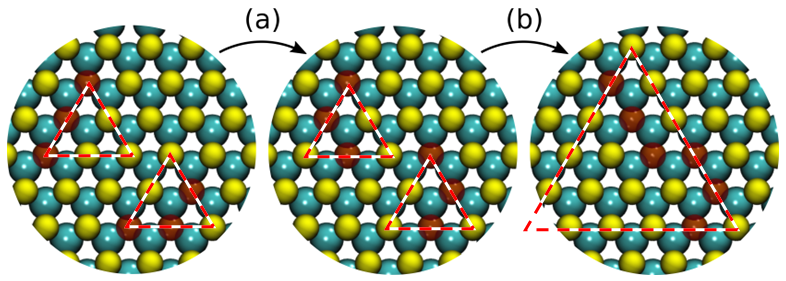}
    \caption{
        Illustration of slow and fast sulfur-vacancy jumps in a top-down view of a MoS\textsubscript{2} monolayer: 
        (a) slow jumps bring the ends of two small vacancy clusters into close proximity; 
        (b) the two small defect clusters associate into a larger one, with the final jump occurring much faster. 
        Sulfur atoms are shown in yellow (matte), sulfur vacancies in red (translucent), and molybdenum atoms in cyan (shiny). 
        The regions accessible to three and six vacancies, respectively, are indicated by dashed red lines.
        The VMD (Visual Molecular Dynamics) software was used to create this figure~\cite{humphrey199633}.
    }
    \label{fig:cluster_capture_scheme}
\end{figure}

The \ac{MLIP} \ac{MD} simulations in Ref.~[\onlinecite{Fltotto2026}] revealed dynamical processes leading to the formation of vacancy clusters arising from the cooperative jump mechanism.
Two or more adjacent vacancies are mobile only within a predefined triangular region because jumps leaving this region are blocked, either by a molybdenum atom beneath the layer or by a neighboring sulfur atom that prohibits the low-energy jump through an interstitial site.
If two vacancy clusters are arranged such that their triangular mobile regions overlap, they can eventually meet and merge into a larger cluster confined to a larger triangular region, as illustrated in Fig.~\ref{fig:cluster_capture_scheme}.
Once merged, clusters typically remain connected due to an energetic stabilization of larger clusters, with only rare dissociation events into smaller sub-clusters driven by entropy.
Accordingly, the association of vacancy clusters and their reverse dissociation into smaller sub-clusters were identified as the key motifs governing large-scale sulfur-vacancy dynamics in MoS\textsubscript{2}~\cite{Fltotto2026}.

Here, we analyze \ac{MLIP} \ac{MD} simulations initialized with individual vacancy clusters to extract rates for two key cooperative vacancy-jump processes: cluster association and cluster dissociation, where operationally a cluster is defined as a set of vacancies occupying neighboring sites in the triangular S-atom layer.
A \ac{KMC} approach that applies these rates to simulate sulfur vacancy dynamics is introduced at the beginning of Sect.~\ref{subsec:clustering}.
The \ac{MD} simulations from which the corresponding rates are extracted apply the \ac{MLIP} that was fine-tuned from the MACE MP-0 foundation model~\cite{Batatia2022, Batatia2023} to best describe vacancy dynamics in MoS\textsubscript{2} and validated in Ref.~\cite{Fltotto2026}.
To access transition rates of the two key processes, we sample the corresponding \acp{FPT} from the \ac{MD} simulations.

\begin{figure}[htbp]
    \centering
    \includegraphics[width=\linewidth]{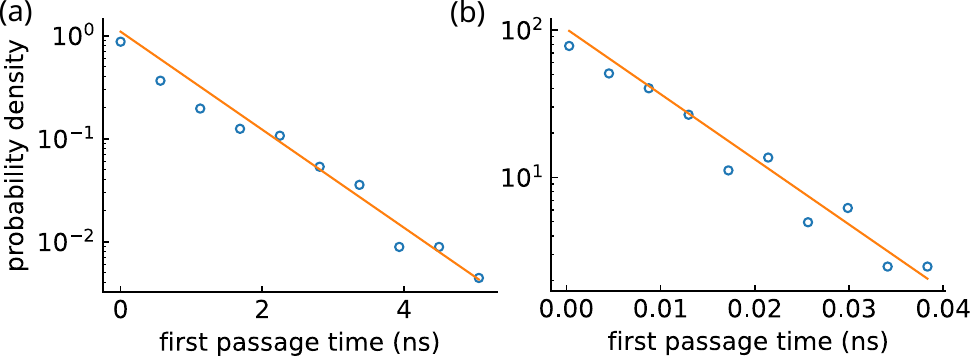}
    \caption{
        Probability density histogram of the \acl{FPT} for transitions between one and two connected sub-clusters: 1$\rightarrow$2 (a) and 2$\rightarrow$1 (b), from \ac{MLIP} \ac{MD} simulations with 3–8 vacancies.
    }
    \label{fig:ftp_fits}
\end{figure}

The analyzed data comprise 390~ns of \ac{MLIP} \ac{MD} simulations distributed over six trajectories containing clusters of 3--8 vacancies at $T = 1000$~K, as discussed at the end of Sect.~\ref{subsec:diffusion_coefs}.
The number of connected sub-clusters resulting from the dissociation of an initially connected ("original") cluster was determined using a depth-first graph traversal algorithm.
To eliminate spurious transitions caused by vibrational motion near interstitial sites, we projected the atomic positions onto a grid of lattice and interstitial sites using a nearest-neighbor assignment.
A transition was recorded only when all sulfur atoms were assigned to lattice sites in the respective frame.
In this way, we determined all transitions from one to two connected clusters (dissociation) and the reverse transitions from two to one connected cluster (association).
Probability density histograms of the corresponding first-passage times $\Delta t$ are shown in Fig.~\ref{fig:ftp_fits}.
Assuming Markovian two-state kinetics, the first-passage-time histograms can be described with the exponential \ac{PDF}
\begin{equation}
    \label{eq:ftp_pdf}
    \rho (\Delta t) = k \exp(-k \, \Delta t)
\end{equation}
with the rate $k$.
Fitting this \ac{PDF} to the \ac{MD} data yields the rates $k_{2 \leftarrow 1}$ for cluster dissociation and $k_{1 \leftarrow 2}$ for cluster association, averaged over cluster sizes and all configurations with the same number of connected sub-clusters.
Notably, the \acp{FPT} extracted from \ac{MD} simulations apparently do not depend on the size of the vacancy cluster or its shape in this regime, as demonstrated in Sect.~1 of the Supplementary Information.
The rates corresponding to the fitted \ac{PDF} shown in Fig.~\ref{fig:ftp_fits} differ by almost two orders of magnitude: 
\begin{eqnarray}
    k_{2 \leftarrow 1} &=& 1.095 \ \mathrm{ns}^{-1} \,=\, \frac{1}{0.913 \ \mathrm{ns}}\label{k12Val}\\
    k_{1 \leftarrow 2} &=& 101.9 \ \mathrm{ns}^{-1} \,=\, \frac{1}{0.0098 \ \mathrm{ns}} \label{k21Val}
\end{eqnarray}

\subsection{Self-organized clustering}
\label{subsec:clustering}
Using these transition rates, we performed \ac{KMC} simulations employing the null-event (rejection) algorithm described in Ref.~[\onlinecite{Chatterjee2007}].
The \ac{KMC} model treats vacancies as the mobile species and includes only cooperative, one-vacancy-assisted jumps, since non-assisted jumps are effectively suppressed by the large energy barrier discussed above~\cite{Fltotto2026}.
Based on the \ac{MLIP} \ac{MD} results (cf. Eqs.~(\ref{k12Val}) and (\ref{k21Val})), the \ac{KMC} model distinguishes between two types of vacancy jumps: a fast process with the rate $k_{1\leftarrow2}$, in which a vacancy jump reduces the number of connected vacancy sub-clusters by one, and a slow process with the rate $k_{2\leftarrow1}$, which comprises all other cooperative vacancy jumps.
These slow jumps either reverse the fast process or preserve the number of connected vacancy clusters, for example by introducing a kink into a previously linear cluster.
The latter cases are treated equally motivated by \ac{DFT} results showing nearly identical energy barriers for these processes~\cite{Gao2021, Wang2022}.
See Sect.~\ref{subsec:methods_kmc} below for more details on the \ac{KMC} implementation.

The \ac{KMC} approach was validated by simulating individual vacancy clusters and comparing the resulting sulfur \acp{MSD} with those obtained from the corresponding \ac{MD} simulations.
The \ac{MSD} is computed for given lag time $\tau$ as average over vacancies $i$ with positions $\mathbf{r}_i (t)$ and over the starting times $t_0$:
\begin{equation}
 \label{eq:msd}
    \mathrm{MSD}(\tau) = \left<\, \left<
    \lvert \mathbf{r}_i(t_0 + \tau) - \mathbf{r}_i(t_0) \rvert^2
    \right>_i \right>_{t_0}.
\end{equation}
As demonstrated in Sect.~2 of the Supplementary Information, both methods yield very similar results, apart from a constant offset arising from the continuous atomic positions in \ac{MD} and the discrete state representation on a lattice in \ac{KMC}.
Here, this offset will be inconsequential, as the diffusion coefficient depends only on the slope of the converged \ac{MSD} according to the \ac{2D} Einstein-Smoluchowski equation:
\begin{equation}
    \label{eq:diffcoef}
    D = \frac{1}{4} \left[ \frac{d}{d\tau} \mathrm{MSD}(\tau) \right]_{\tau \rightarrow \infty}
\end{equation}

As a starting point for investigating long-range migration in this \ac{KMC} model, we analyze three 10\,\textmu s simulations with sulfur vacancy concentrations $c = {N_\mathrm{vac}}/{N_\mathrm{sites}}$ of 7, 12, and 18\,\% in a single sulfur layer with random initial vacancy positions in a supercell with $70 \times 70$ sulfur sites. The convergence of the supercell size is demonstrated in Sect.~3 of the Supplementary Information.

Figure~\ref{fig:jumpvis} shows the number of vacancy transitions between all pairs of lattice sites, clearly revealing mobile regions arising from cooperative vacancy jumps within clusters as also predicted by \ac{MD} simulations~\cite{Fltotto2026}.
From these transition counts, we determined an adjacency matrix of all lattice sites and counted the number of connected clusters using a depth-first search.
Figure~S4 in the Supplementary Information additionally shows the time evolution of the corresponding density of lattice site clusters that are connected by vacancy jumps in these three \ac{KMC} simulations.
Three different vacancy concentrations are considered, each exhibiting distinct clustering behavior:
\begin{figure*}[htbp]
    \centering
    \includegraphics[width=\linewidth]{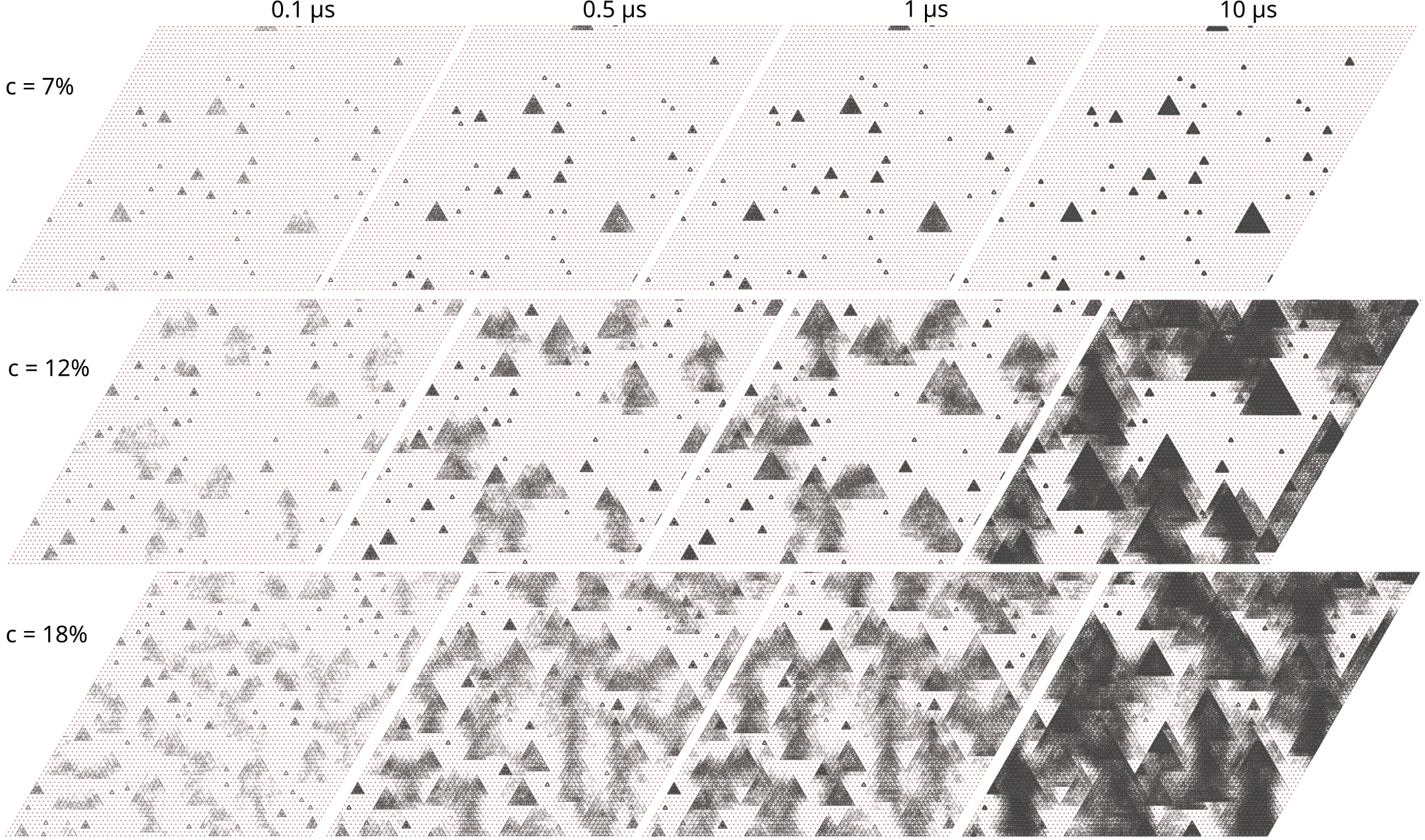}
    \caption{
        Time evolution of the number of sulfur-vacancy transitions between lattice site pairs from \ac{KMC} simulations with vacancy concentrations of 7\,\%, 12\,\%, and 18\,\%.
        Transitions were sampled cumulatively from the beginning of the simulation up to 0.1~\textmu s, 0.5~\textmu s, 1~\textmu s and 10~\textmu s at intervals of 1~ns.
        Lattice sites are implied by small spheres, and each pair that was connected by a transition is indicated by a gray cylinder, with diameter scaled logarithmically to the transition count.
        Transitions across periodic boundaries are not shown, although periodic boundary conditions were applied during the simulations.
    }
    \label{fig:jumpvis}
\end{figure*}

(\textit{i}) At a low vacancy concentration of 7\,\%, only small clusters form in regions with a local vacancy density above the average. During the 10\,\textmu s simulation shown in the top row of Fig.~\ref{fig:jumpvis}, these clusters explore their entire triangular mobile regions.
However, the mobile regions of different clusters rarely overlap, except at the very beginning of the simulation.
Consequently, the clusters remain spatially localized.
This results in a density of connected lattice sites clusters that is constant in time as shown in Fig.~S4 of the Supplementary Information.

(\textit{ii}) At an intermediate vacancy concentration of 12\,\% (center row of Fig.\ref{fig:jumpvis}), clusters with overlapping mobile regions become evident within the first few hundred nanoseconds of the simulation.
As the simulation progresses, these clusters merge, expanding their triangular mobile regions.
After 10\,\textmu s, many of the larger clusters have explored their entire mobile regions, which now overlap with additional clusters.
Further expansion is kinetically hindered on these timescales: vacancies in two large clusters with overlapping mobile regions must arrive simultaneously at the intersection of their mobile regions for both clusters to associate.
Although diffusion across the entire supercell would be possible, this kinetic constraint leaves some areas of the supercell unexplored, at least for all relevant time scales.

(\textit{iii}) At a high vacancy concentration of 18\,\%, many overlapping clusters form rapidly, creating a connected network that is progressively explored.
Accordingly, the density of connected lattice site clusters decreases rapidly (Fig.~S4 of the Supplementary Information).
After 10\,\textmu s, only relatively small regions remain unexplored, and numerous diffusion pathways span the entire supercell.

\begin{figure*}[htbp]
    \centering
    \includegraphics[width=0.83\linewidth]{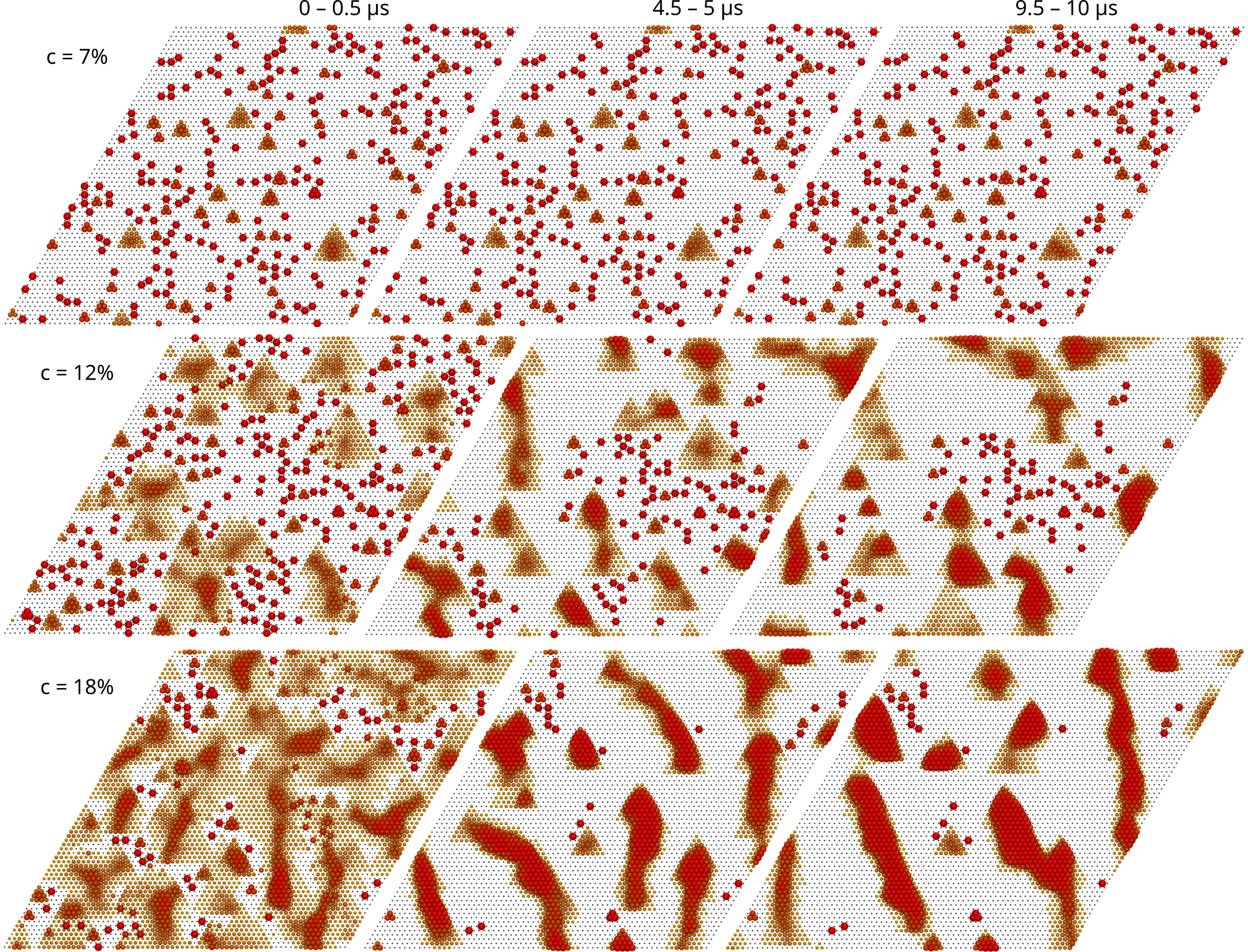}
    \caption{
        Site-resolved sulfur vacancy occupation function averaged within 0.5~\textmu s intervals shown in the complete $70 \times 70$ S sites supercell from \ac{KMC} simulations at concentrations of 7~\%, 12~\%, and 18~\%.
        Lattice sites that were vacant during the respective time interval are represented by spheres, with colors ranging linearly from yellow (low occupation probability) to red (high occupation probability) and the radius increasing linearly with the occupation probability.
        Lattice sites that were not vacant are indicated as small, gray dots.
    }
    \label{fig:occupationvis}
\end{figure*}

These observations are illustrated from a complementary perspective by Fig.~\ref{fig:occupationvis}: It presents the site-resolved vacancy occupation function, averaged over several 0.5\,\textmu s time intervals for the same simulations as shown in Fig.~\ref{fig:jumpvis}.
At a vacancy concentration of 7~\%, only relatively small clusters form, even after 10~\textmu s. 
In addition, large regions containing isolated, essentially immobile vacancies persist. 
In the intermediate regime at 12~\%, after 10~\textmu s, some regions are completely free of stationary vacancies due to the formation of larger clusters, which are mobile as a result of cooperative vacancy jumps. 
However, other regions of the simulation cell still contain stationary vacancies or small clusters with restricted mobility. 
At high vacancy concentrations (here $c = 18~\%$), only a few but very large vacancy clusters form. 
Due to their effectively unrestricted mobility, most vacancies that were initially isolated are incorporated into clusters after 10~\textmu s.

While the transition counts (Fig.~\ref{fig:jumpvis}), with their triangular patterns resulting from cooperative jumps, highlight the mechanism underlying vacancy cluster formation and migration, the averaged vacancy occupations (Fig.~\ref{fig:occupationvis}) provide insight into the cluster morphology. At any given time during the simulations, the larger clusters exhibit a rather elongated shape within their triangular accessible region, oriented along one of the symmetry directions.
These shapes exhibit well-defined boundaries along threefold axes of the hexagonal lattice, which arise from the selection rules governing cooperative vacancy jumps.

Videos with animated trajectories of the three corresponding \ac{KMC} simulations are provided in Ref.~[\onlinecite{refodat_mos2_kmc}].
At intermediate (12~\%) and high (18~\%) vacancy concentrations, vacancy clusters partially agglomerate into large voids. 
Although the formation of such voids leads to an increase in mechanical stress in the monolayer~\cite{Komsa2013}, it has previously been demonstrated in \ac{TEM} measurements at $T = 800^\circ\mathrm{C}$ following high-energy electron irradiation~\cite{Chen2018}.
While such large clusters are present at both intermediate and high concentrations in our simulations, small vacancy clusters and isolated, stationary vacancies clearly persist at a vacancy concentration of 12~\% in our \ac{KMC} simulations, whereas only few remain after equilibration at 18~\%.

To further investigate the clustering behavior, we define a vacancy structure factor $S(\mathbf{k})$ via the discrete Fourier transform of the zero-centered vacancy occupation function $n(\mathbf{R})$ in \ac{2D}.
\begin{align}
    n(\mathbf{R}, t) &\,=\, \delta_{\mathrm{vac}}(\mathbf{R}, t) - c \\
    S(\mathbf{k}, t) &\,=\,  \textstyle{\frac{1}{N}\,\big|\, \sum_{\mathbf{R}} n(\mathbf{R}, t) \, e^{-i \mathbf{k} \cdot \mathbf{R}} \,\big|^{\,2}}
    \label{eq:structure_factor}
\end{align}
Here, $c = N_\mathrm{vac} / N$ denotes the vacancy concentration and $\delta_{\mathrm{vac}}(\mathbf{R}, t) = 1$ if, at a given time $t$, the lattice site $\mathbf R$ is vacant and 0 else.

\begin{figure}[htbp]
    \centering
    \includegraphics[width=\linewidth]{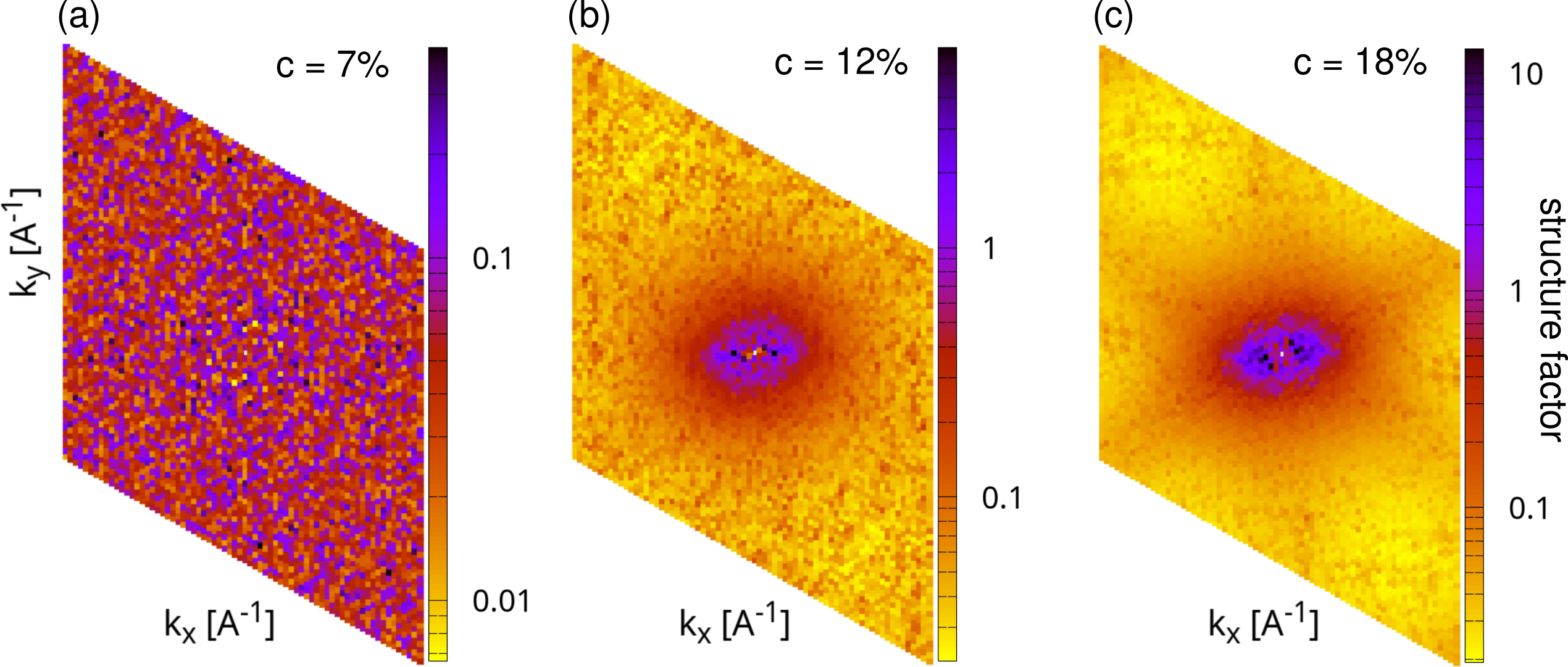}
    \caption{
        Vacancy structure factor computed via Eq.~\ref{eq:structure_factor} from the same \ac{KMC} simulations as Fig.~\ref{fig:jumpvis} and \ref{fig:occupationvis}.
        After an initial equilibration period of 1\,\textmu s, the data were averaged over the subsequent 9\,\textmu s of simulation time.
        The diagrams show the top view on the \ac{2D} reciprocal cell, centered around the $\Gamma$-point.
        The structure factor at the $\Gamma$-point is zero by definition and has been omitted here.
    }
    \label{fig:2dft}
\end{figure}

The resulting structure factors averaged over 9\,$\mu$s are shown in Fig.~\ref{fig:2dft}.
At a low vacancy concentration of 7\,\%, the structure factor is essentially featureless and dominated by noise, indicating that the corresponding real-space vacancy distribution remains largely disordered with little to no clustering and no longe-range correlations.
At this concentration, large regions of the simulation cell exhibit no mobile vacancies (Figure~\ref{fig:jumpvis}).
Instead, vacancies in these regions stationary and therefore preserve their initial random distribution.

A different picture emerges at \(c = 12\,\%\).
As already suggested by the vacancy transition counts in Fig.~\ref{fig:jumpvis}, vacancies begin to aggregate into finite-sized clusters.
This behavior is reflected in the structure factor by the appearance of a broad peak around the \(\Gamma\)-point, accompanied by a largely featureless background at larger \(|\mathbf{k}|\).

At a vacancy concentration of 18\,\%, the structure factor changes qualitatively once more.
While the broad peak at the \(\Gamma\)-point persists—indicating continued clustering—the structure factor now exhibits pronounced lines along the high-symmetry directions defined by the reciprocal lattice vectors \(\mathbf{b}_1\) and \(\mathbf{b}_2\) as well as \(\mathbf{b}_1 + \mathbf{b}_2\).
This reveals that the vacancy clusters in real-space are highly anisotropic, thereby affirming and quantifying the visual impression of the real-space vacancy occupation shown in Fig.~\ref{fig:occupationvis}.

Overall, the structure factors shown in Fig.~\ref{fig:2dft} highlight clear qualitative differences in the vacancy dynamics across the three concentrations, ranging from a disordered regime with small, localized clusters and localized vacancies to the formation of larger clusters and, finally, to anisotropically extended clusters.

In the following, the influence of the sulfur-vacancy concentration on vacancy dynamics is investigated systematically using \ac{KMC} simulations with concentrations ranging from 5\,\% to 20\,\% in steps of 1\,\%.
Such concentrations have been shown to be experimentally achievable using a variety of techniques~\cite{Ma2013, Cheng2016, Bertolazzi2017, Tsai2017, Li2019, Lee2021, Man2024}.
However, a direct quantitative comparison with experimentally reported vacancy concentrations is often challenging, since the simulation values here strictly refer to the vacancy concentration within a single sulfur layer, whereas some experimental techniques are not exclusively sensitive to the topmost layer.
A notable exception is vacancy quantification using scanning probe methods~\cite{Lee2021}.

For each vacancy concentration analyzed here, ten simulations were performed with different random initial vacancy positions to capture the influence of structural disorder on vacancy dynamics. 
Vacancy jumps between lattice sites were sampled at intervals of 2~ns to construct an adjacency matrix, from which the number of connected clusters of sulfur sites was determined. 
After an initial equilibration period of 4.5\,\textmu s in the \ac{KMC} simulations, the density of connected lattice-site clusters was evaluated over the subsequent 0.5\,\textmu s of each simulation. 

\begin{figure}[htbp]
    \centering
    \includegraphics{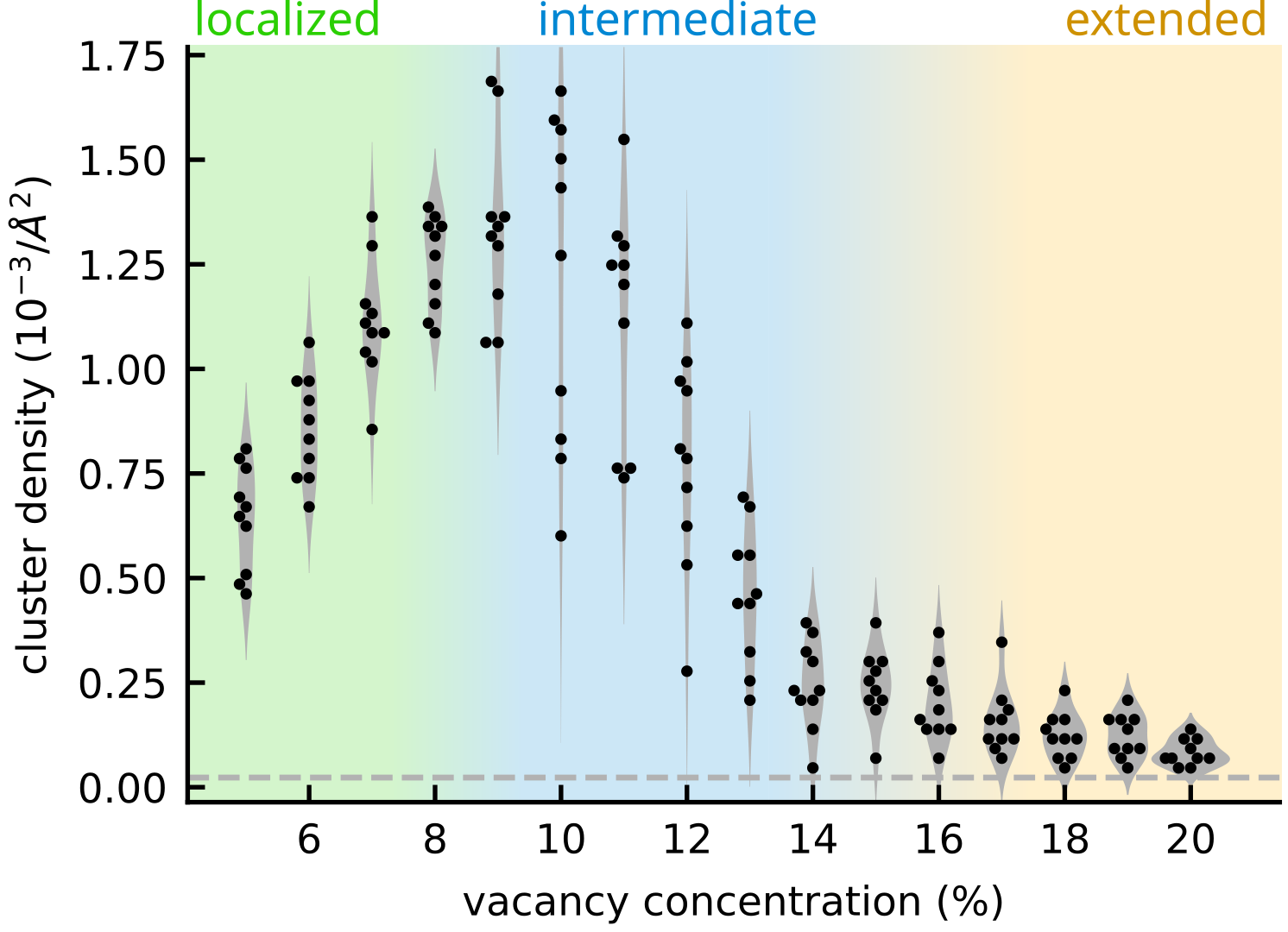}
    \caption{
        Density of connected vacancy clusters as a function of vacancy concentration.
        For each vacancy concentration, ten \ac{KMC} simulations of 5\,\textmu s each were performed.
        The simulations were initialized with random vacancy positions and have been equilibrated for 4.5\,\textmu s before determining the lattice site pairs connected by vacancy jumps during the final 0.5\,\textmu s of each simulation.
        Data points were horizontally shifted for clarity.
        Gray violin plots indicate kernel densities at each concentration.
        The calculations were performed in a supercell of $70 \times 70$ sulfur-lattice sites.
        The minimal density, which corresponds to the inverse of the supercell area, is indicated by a dashed horizontal line.
    }
    \label{fig:cluster_counts}
\end{figure}

The resulting dependence of the density of connected vacancy clusters on the vacancy concentration is shown in Fig.~\ref{fig:cluster_counts}. 
As illustrated already by representative simulations in Fig.~\ref{fig:jumpvis} and \ref{fig:2dft}, three distinct regimes emerge with increasing vacancy concentration.

In the first regime, the density of connected lattice-site clusters increases trivially as additional neighboring vacancies appear and form small clusters.
Since the corresponding mobile regions do not overlap, these clusters remain localized.
Hence, we refer to this regime as the localized regime.
The density of lattice site clusters reaches a maximum at a vacancy concentration of about 9\,\%.
In an intermediate regime, mobile regions of vacancy clusters begin to overlap. This overlap promotes the association of clusters, resulting in larger clusters that progressively form a connected network in a self-organized manner resulting in a decrease of the number of independent clusters.
Beside these larger clusters, numerous small connected lattice site clusters are still visible in this regime, hinting towards a critical, scale-free behavior at these intermediate vacancy concentrations.
These small clusters typically show larger jump rates between the few sites and therefore appear darker in Fig.~\ref{fig:jumpvis}.
Finally, at a vacancy concentration of approximately 14\,\%, the density of connected lattice site clusters begins to saturate, eventually reaching a single cluster in three of the ten simulations at a vacancy concentration of 20\,\%.
In this regime, the clusters extend over large portions or even the entirety of the $70 \times 70$ site supercell (cf. bottom row of Fig.~\ref{fig:jumpvis}).

\subsection{Vacancy diffusion coefficients}
\label{subsec:diffusion_coefs}
\begin{figure}[htbp]
    \centering
    \includegraphics{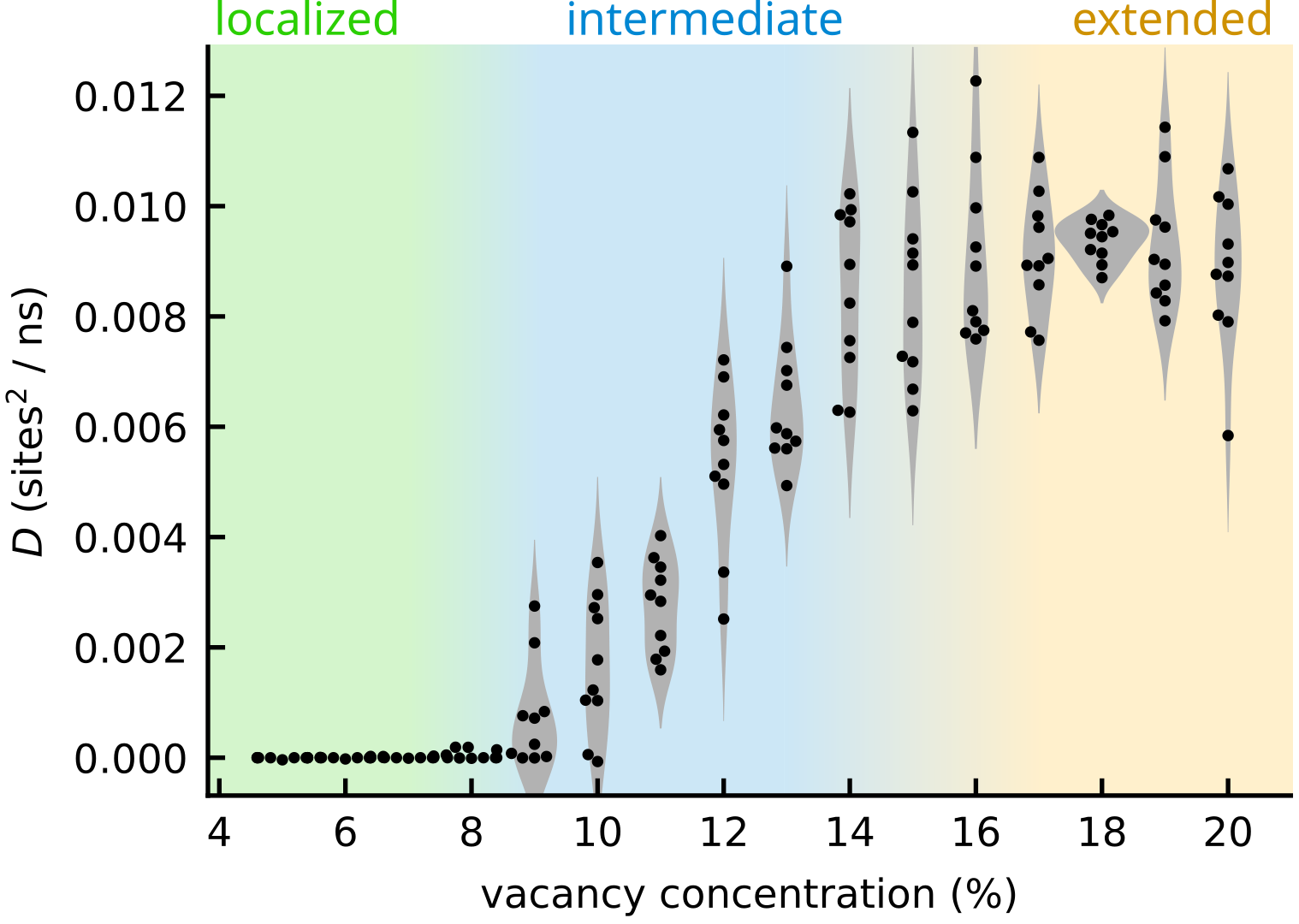}
    \caption{
        Diffusion coefficient $D$ of sulfur vacancies as a function of concentration.
        Results were obtained from the same ten \ac{KMC} simulations for each concentration as analyzed in Fig.~\ref{fig:cluster_counts}.
        Diffusion coefficients were determined using a linear regression to the linear regime of the \ac{MSD} (Eq.~\ref{eq:diffcoef}).
    }
    \label{fig:diffcoefs}
\end{figure}

After identifying the three regimes of concentration dependent sulfur-vacancy mobility in MoS\textsubscript{2}, we determine the corresponding diffusion coefficients.
The vacancy \ac{MSD} was calculated according to Eq.~\ref{eq:msd} from the same \ac{KMC} simulations discussed for Fig.~\ref{fig:cluster_counts}.
As shown in Sect.~5 of the Supplementary Information, the \ac{MSD} exhibits linear lag-time dependence within the simulated time window, indicating diffusive behavior.
Diffusion coefficients $D$ were obtained from linear regression over the interval  $\tau \in \left[2.5 \ \text{\textmu s}, 3.5 \ \text{\textmu s} \right]$.
As shown in Fig.~\ref{fig:diffcoefs}, the three distinct regimes identified from the analysis of connected lattice-site clusters are clearly observed in the concentration-dependent diffusion coefficient as well.

In the localized regime, the vacancy \ac{MSD} initially increases (shown in Sect.~5 of the Supplementary Information).
However, after a characteristic lag time required for vacancies in the numerous small clusters to explore their full (typically very small) mobile regions, the \ac{MSD} reaches a plateau.
As a result, no long-range diffusion is observed.
The diffusion coefficient begins to increase from zero at a vacancy concentration of approximately 9\,\%, which conicides within the numerical accuracy with the maximum in the density of connected lattice-site clusters in Fig.~\ref{fig:cluster_counts}.
The diffusion coefficient increases with vacancy concentration as more clusters associate, thereby enlarging their mobile regions.
At the transition between the intermediate and extended regimes, where the density of connected lattice-site clusters begins to saturate, the diffusion coefficient likewise levels off.
At vacancy concentrations above $\sim 14$\,\%, the diffusion coefficient approaches an approximately constant value and remains nearly unchanged up to the highest analyzed vacancy concentration of 20\,\%.

Finally, we make two brief remarks:

(\textit{i}) The emergence of three identical vacancy-concentration regimes both for the  density of connected lattice-site clusters and for the diffusion coefficient (Fig.~\ref{fig:cluster_counts} and \ref{fig:diffcoefs}) indicates the presence of two distinct mechanisms suppressing vacancy diffusion:
At low and intermediate concentrations ($c\lesssim 14\%$), the diffusion coefficient is governed by the size of individual vacancy clusters, whose mobility is confined to spatially limited regions until they merge with other clusters.
In contrast, extended clusters form at high concentrations ($c\gtrsim 14\%$) and the diffusion coefficient $D$ becomes limited by the rate of collective migration within their mobile regions; consequently, further increases in cluster size do not lead to higher diffusion coefficients, yielding approximately constant $D$, see Fig.~\ref{fig:diffcoefs}.

(\textit{ii}) 
The rates used in these \ac{KMC} simulations (Eq.~(\ref{k12Val},\ref{k21Val})) are derived from \ac{MD} simulations of MoS\textsubscript{2} performed at a high temperature (1000~K), which currently is of considerable interest due to the demand of electronic components for spezialized, high-temperature applications~\cite{Wu2026}.
Determining diffusion coefficients for other, lower temperatures naturally requires additional modeling, but the emergence of the three distinct concentration regimes is expected to depend only weakly on temperature.


\subsection{Summary}
We analyzed \ac{MLIP} simulations of individual sulfur-vacancy clusters in MoS\textsubscript{2} at $T=1000$ K, spanning hundreds of nanoseconds.
Averaging over cluster shapes and sizes, we extracted rates for two processes previously identified as key mechanisms governing large-scale sulfur-vacancy dynamics.
These are the dissociation of a connected cluster into two sub-clusters and the much faster reverse association process.
These processes were used to parameterize a \ac{KMC} model.

Using the \ac{KMC} model, we computed diffusion coefficients in large supercells with initially random vacancy distributions and resolved their dependence on concentration.
Three regimes emerge:
(i) A localize regime showing only small, non-overlapping clusters resulting in the absence of long-range diffusion.
(ii) In an intermediate regime, the diffusion coefficient increases with vacancy concentration due to enhanced, self-organized association of initially distinct clusters.
(iii) Above concentrations of $\sim$14\%, extended clusters are formed constituting a connected network, and the diffusion coefficient approaches a plateau.
In particular, the large vacancy clusters formed in this regime are extended anisotropically as shown by the \ac{2D} spatial Fourier transform of the vacancy occupation function.

These findings bridge the gap between atomistic \ac{MD} simulations, which resolve the underlying mechanisms, and continuum transport models of MoS\textsubscript{2}-based devices.
Together with the previously established link between sulfur-vacancy mobility and memristive behavior in MoS\textsubscript{2}~\cite{Li2018,Spetzler2024}, the concentration dependence identified here provides a key insight into the microscopic origin of this effect.
This understanding offers new opportunities for the targeted control of defect dynamics and the optimization of MoS\textsubscript{2}-based memristive devices.

\section{Methods}
\subsection{Machine learning interatomic potential molecular dynamics}
\label{subsec:methods_mlipmd}
The \ac{MLIP} employed in this work was fine-tuned from the MACE MP-0 foundation model~\cite{Batatia2022,Batatia2023} using data of MoS\textsubscript{2} monolayers with two adjacent sulfur vacancies, as described and extensively validated in our previous work~\cite{Fltotto2026}. 
The training and test data were generated using the CP2K software package~\cite{Khne2020}, following the procedure outlined in Ref.~[\onlinecite{Fltotto2026}]. 
The resulting \ac{MLIP} model, together with the corresponding training and test data, is openly available~\cite{Floetotto2025MLIPMoS2Data}.

The \ac{MLIP}-based \ac{MD} simulations were performed for MoS\textsubscript{2} monolayers containing three to eight adjacent vacancies using the ASE software package~\cite{HjorthLarsen2017}. 
The simulations were carried out with a time step of 0.5~fs at a temperature of 1000~K, employing a Langevin thermostat.

\subsection{Kinetic Monte-Carlo}
\label{subsec:methods_kmc}
The \ac{KMC} simulations reported in this work were performed using the rejection \ac{KMC} algorithm described in Ref.~[\onlinecite{Chatterjee2007}].
The source code is available in Ref.~[\onlinecite{mos2_montecarlo}].
Each \ac{KMC} step begins with the random selection of a vacancy for a potential jump. 
Subsequently, all possible jumps of that vacancy that conform to the vacancy-assisted jump mechanism characteristic of MoS\textsubscript{2}, as described in the main text, are identified and the corresponding rates are enumerated. 
These rates are normalized by the maximal total jump rate for a single vacancy, resulting in a total rate (the sum of all rates at the site) in the range $[0,1]$. 
A uniform random number in the same range is then drawn. 
If the random number exceeds the total rate, the attempted jump is rejected. 
Otherwise, a specific jump is selected in the typical \ac{KMC} manner via the position of the random number within the cumulative sum of the individual jump rates.

With the \ac{MD}-sampled jump rates described in the context of Fig.~2 in the main manuscript, the maximal total rate $k_\mathrm{max}$ for a single vacancy equals the rate of cluster association. 
As proposed in Ref.~[\onlinecite{Chatterjee2007}], the simulation time is advanced in each \ac{KMC} iteration by the average increment $1 / \left( N k_\mathrm{max} \right)$, where $N$ is the number of vacancies in the system. 
Using an averaged time increment is justified because the results reported in this work depend only on dynamic quantities involving a large number of \ac{KMC} events.

\section*{Data availability}
The fine-tuned MACE \ac{MLIP} model, together with the corresponding fine-tuning data as well as the \ac{MLIP} training and \ac{DFT} input files, are publicly available in Ref.~[\onlinecite{Floetotto2025MLIPMoS2Data}].
The \ac{KMC} trajectories underlying Fig.~\ref{fig:jumpvis}, Fig.\ref{fig:occupationvis} and Fig.~\ref{fig:2dft} here as well as Fig.~S4 in the Supplementary Information together with corresponding animations are published in Ref.~[\onlinecite{refodat_mos2_kmc}].
The remaining data are available from the authors upon request.

\section*{Code availability}
The source code used to run \ac{KMC} simulations is available in Ref.~[\onlinecite{mos2_montecarlo}].
The source code used to analyze \ac{MD} and \ac{KMC} trajectories is available in Ref.~[\onlinecite{spam_github}].

\section*{Acknowledgements}
The authors thank Henning Schwanbeck and his colleagues at the University Computing Center of the TU Ilmenau for excellent working conditions and continued support.
In early stages of this work support through the projects “MemWerk” and “Ilmenau School of Green Electronics” (P2022-00-135) that were made possible by funding from the
Carl–Zeiss–Stiftung was received and is gratefully acknowledged.

\section*{Author contributions}
All authors contributed equally to the conceptualization of the study.
E.R. and C.D. supervised the work throughout the project.
A.F. developed the kinetic Monte-Carlo software, carried out the investigation, performed the formal analysis, and prepared the visualizations.
A.F. wrote the original draft of the manuscript.
All authors reviewed and edited the manuscript.

\section*{Competing interest}
The authors declare no competing interests.

\bibliography{exportNoAbstract}

\end{document}



\title{
    Supplementary Material:\\[3mm]
Self-organized defect clustering and\\ concentration-dependent vacancy diffusion in MoS\textsubscript{2}
}

\author{Aaron Flötotto}
\email{aaron.floetotto@tu-ilmenau.de}
\affiliation{%
  Technische Universität Ilmenau, Institute of Physics,\\
  Weimarer Straße 25, 98693 Ilmenau, Germany
}%
\affiliation{
  Institute of Micro- and Nanotechnologies,\\
  Ehrenbergstraße 29, 98693 Ilmenau, Germany
}

\author{Benjamin Spetzler}
\affiliation{
  Christian-Albrechts-Universität zu Kiel, Institute of Materials Science,\\
  Kaiserstraße 2, 24143 Kiel, Germany
}%

\author{Martin Ziegler}
\affiliation{
  Christian-Albrechts-Universität zu Kiel, Institute of Materials Science,\\
  Kaiserstraße 2, 24143 Kiel, Germany
}%

\author{Erich Runge}
\affiliation{%
  Technische Universität Ilmenau, Institute of Physics,\\
  Weimarer Straße 25, 98693 Ilmenau, Germany
}%
\affiliation{
  Institute of Micro- and Nanotechnologies,\\
  Ehrenbergstraße 29, 98693 Ilmenau, Germany
}

\author{Christian Dreßler}
\affiliation{%
  Technische Universität Ilmenau, Institute of Physics,\\
  Weimarer Straße 25, 98693 Ilmenau, Germany
}%
\affiliation{
  Institute of Micro- and Nanotechnologies,\\
  Ehrenbergstraße 29, 98693 Ilmenau, Germany
}

\date{\today}


\maketitle

\acrodef{MD}{molecular dynamics}
\acrodef{DFT}{density functional theory}
\acrodef{NEB}{nudged elastic band}
\acrodef{MLIP}{machine learning interatomic potential}
\acrodefplural{MLIP}[MLIPs]{machine learning interatomic potentials}
\acrodef{RDF}{radial distribution function}
\acrodef{MSD}{mean squared displacement}
\acrodefplural{MSD}[MSDs]{mean squared displacements}
\acrodef{AIMD}{\textit{ab initio} molecular dynamics}
\acrodef{GAP}{Gaussian approximation potential}
\acrodefplural{GAP}[GAPs]{Gaussian approximation potentials}
\acrodef{GPR}{Gaussian process regression}
\acrodef{ACE}{atomic cluster expansion}
\acrodef{GNN}{graph neural network}
\acrodef{SCF}{self-consistent field}
\acrodef{KMC}{kinetic Monte-Carlo}
\acrodef{MC}{Monte-Carlo}
\acrodef{TEM}{transmission electron microscopy}
\acrodef{FPT}{first-passage time}
\acrodefplural{FPT}[FPTs]{first-passage times}

\acresetall 
\vspace*{4cm}

\hspace*{2.5cm}
\parbox{12cm}{
The Supplementary Material addresses the following issues:
\begin{enumerate}[label=\Roman*]
    \item Cluster-size dependence of first-passage times
    \item Comparison between \ac{KMC} and \ac{MLIP} \ac{MD} simulations
    \item Cell size convergence for \ac{KMC} simulations
    \item Equilibration of the density of connected lattice site clusters
    \item Diffusivity from sulfur vacancy mean squared displacements
    \item Links to additional material
\end{enumerate}
}
\newpage
\FloatBarrier
\section{Cluster-size dependence of first-passage times}

\begin{figure}[htbp]
    \centering
    \includegraphics[width=0.95\linewidth]{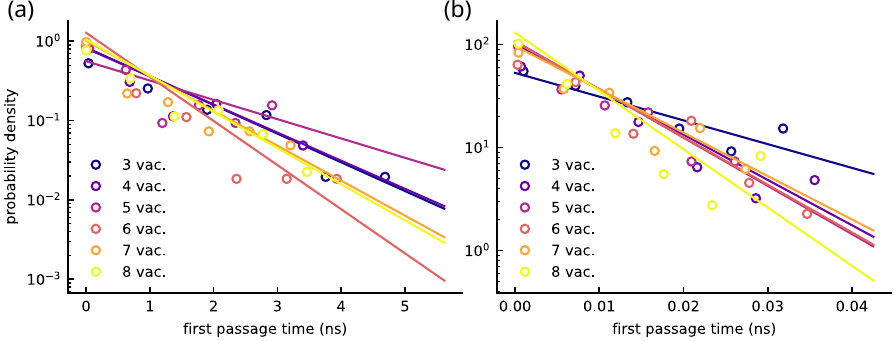}
    \caption{
        Probability density histogram for the \ac{FPT} of the transition from 1 to 2 (a) and 2 to 1 (b) connected sub-clusters in \ac{MLIP} \ac{MD} simulations with 3-8 vacancies.
        In contrast to Figure 2 of the main article, separate probability densities and (exponential) fits are shown for each simulation and, therefore, each cluster size.
    }
    \label{fig:ftp_fits_resolvedclustersize}
\end{figure}

As described in the main manuscript, the rates for dissociation and association of vacancy clusters in \ac{KMC} simulations were determined by evaluating corresponding \acp{FPT} in \ac{MLIP}-based \ac{MD} simulations. 
The \acp{FPT} shown in Fig.~2 of the main manuscript were sampled from multiple \ac{MD} simulations of clusters containing 3 to 8 vacancies, comprising a total simulation time of 390~ns. 
To assess the validity of averaging \acp{FPT} over different cluster sizes, we fitted the \acp{FPT} data from each simulation individually by exponential curves, i.e. straight lines on logarithmic plots. 
As shown in Fig.~\ref{fig:ftp_fits_resolvedclustersize}, the resulting probability density curves exhibit significant scatter due to the reduced data quantity in each individual curve.
Importantly, no systematic trend is observed linking cluster size to the \ac{FPT} for either cluster dissociation or association.

\FloatBarrier
\section{Comparison between \ac{KMC} and \ac{MLIP} \ac{MD} simulations}

\begin{figure}[htbp]
    \centering
    \includegraphics[width=0.95\linewidth]{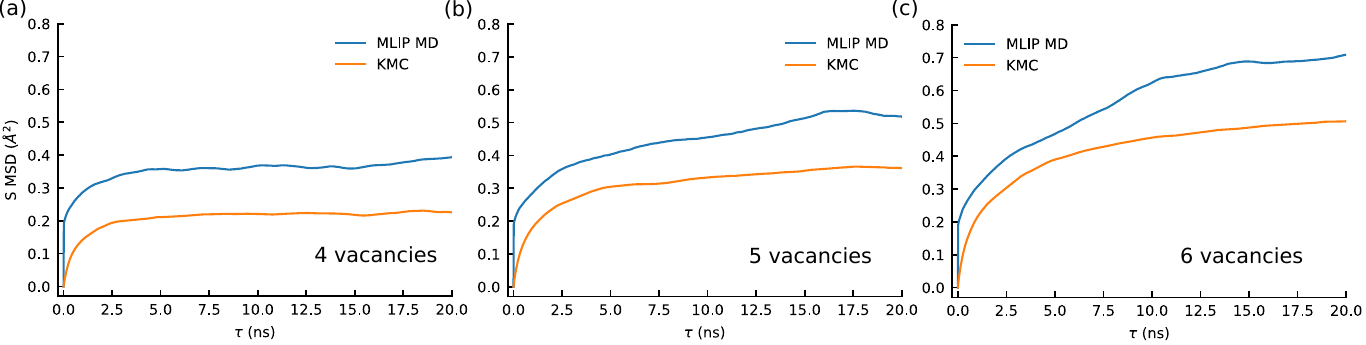}
    \caption{
        Comparison between \acp{MSD} calculated from \ac{MLIP} \ac{MD} simulations and \ac{KMC} simulations.
        The comparison is shown for simulations of one individual cluster --- starting as a straight line of sulfur vacancies.
        The clusters contain 4~(a), 5~(b) or 6~(c) vacancies, respectively.
    }
    \label{fig:msdcomp}
\end{figure}

In order to check the validity of the assumptions made during the construction of the \ac{KMC} model, we compared the resulting sulfur atom dynamics with reference data from \ac{MLIP}-based \ac{MD} simulations.
Figure~\ref{fig:msdcomp} shows a comparison of sulfur atom \acp{MSD} (Eq.~4 of the main manuscript) from \ac{MD} and \ac{KMC} simulations of an individual vacancy cluster consisting of 4, 5 and 6 vacancies, respectively.

The \ac{MSD} curves from \ac{MD} simulations show a rapid initial increase due to atomic vibrations around the equilibrium position, which obviously is not present in the \ac{KMC} simulations based on discrete positions.
This yields an approximately constant offset between the corresponding \ac{MSD} curves.
Except for this constant offset, the two methods yield very similar results.
For the small cluster consisting of 4 vacancies, both methods predict the sulfur \ac{MSD} to reach a constant value after a lag time of about 4~ns.
For the larger vacancy clusters consisting of 5 and 6 vacancies, the two methods result in very similar slopes of the \ac{MSD} curves.
Therefore, the \ac{KMC} model is also able to estimate sufficiently accurate diffusion coefficients (Eq.~5 of the main manuscript) for these defect clusters.

It is worth noting that the time required for a cluster to sufficiently explore its mobile region increases with cluster size. 
This limits the cluster sizes for which converged \ac{MLIP}-based \ac{MD} simulations can be obtained.

\FloatBarrier
\section{Cell size convergence for \ac{KMC} simulations}

\begin{figure}[htbp]
    \centering
    \includegraphics[width=0.5\linewidth]{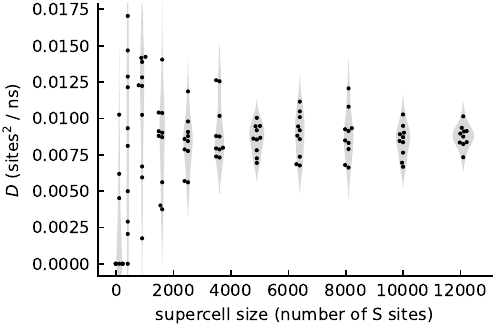}
    \caption{
        Convergence test for the supercell size with random initial vacancy positions and a vacancy density of 14~\% showing the supercell size dependence.
        10 3~$\mu s$ long KMC simulations were performed for each supercell size.
        Diffusion coefficients were calculated from a linear regression to the linear part of the \ac{MSD} according to Equation~5 of the main article.
        Later \ac{KMC} simulations were performed with a supercell size of 4900 sulfur sites leading to a supercell edge length of 22.3~nm.
    }
    \label{fig:cellconv}
\end{figure}

Finite-size effects can significantly influence Monte Carlo simulations of cooperative vacancy hopping, as small supercell sizes may lead to a dependence of dynamical quantities such as the \ac{MSD} and diffusion coefficient $D$ on the initial configuration~\cite{Jckle1994}. 
To ensure that such effects do not affect the results reported in this work, we performed a convergence test of the vacancy diffusion coefficient with respect to the supercell size. 
These simulations were carried out at an intermediate vacancy concentration of 14~\%. 
The diffusion coefficients were obtained from the slope of the \ac{MSD} curves at long lag times, according to Eq.~5 in the main manuscript. 
As shown in Fig.~\ref{fig:cellconv}, the diffusion coefficient converges and exhibits only relatively little variation between simulations with different initial vacancy configurations at a supercell size of 4900 sulfur sites, corresponding to a $70 \times 70$ cell, which was therefore used in all subsequent simulations. 
This variability between simulations with identical parameters but different initial void configurations is also evident in the presentation of the key results Figs.~4 and 5 of the main manuscript.

\FloatBarrier
\section{Equilibration of the density of connected lattice site clusters}

The number of clusters of lattice sites connected by vacancy jumps, identified within a given time interval of a \ac{KMC} simulation, evolves over the course of the simulation. 
The simulations reported here are initialized with a random distribution of vacancies across the supercell. 
During the simulation, vacancies migrate exclusively via the cooperative jump mechanism described previously~\cite{Fltotto2026}, leading to self-organization into clusters over time.

\begin{figure}[htbp]
    \centering
    \includegraphics[width=0.5\linewidth]{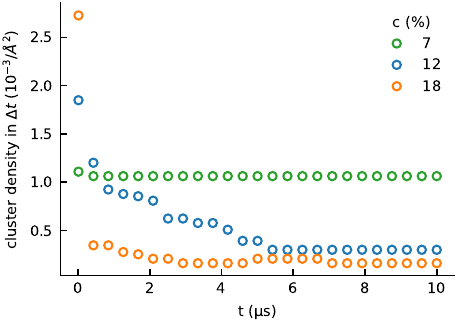}
    \caption{
        Time evolution of the density of connected clusters among lattice sites that participated in vacancy jumps for different vacancy concentrations $c$, corresponding to the simulations visualized in Fig.~3 of the main article. 
    Each data point was obtained by identifying connected lattice-site pairs within the time interval since the previous data point or since the beginning of the simulation. 
    Clusters among these connected sites were identified using a depth-first graph traversal algorithm. 
    Only lattice sites that were occupied by a vacancy at least once during the simulation were considered.
    }
    \label{fig:cluster_counts}
\end{figure}

Figure~\ref{fig:cluster_counts} shows the equilibration of the density of connected lattice-site clusters over different time intervals in the \ac{KMC} simulations, which were also analyzed in Fig.~3 of the main manuscript. 
The three vacancy concentrations considered here lie within the three distinct diffusion regimes identified in this work. 
In all three cases, the equilibration is reached at approximately 5~ns.

\FloatBarrier
\section{Diffusivity from sulfur vacancy mean squared displacements}

\begin{figure}[htbp]
    \centering
    \includegraphics[width=0.9\linewidth]{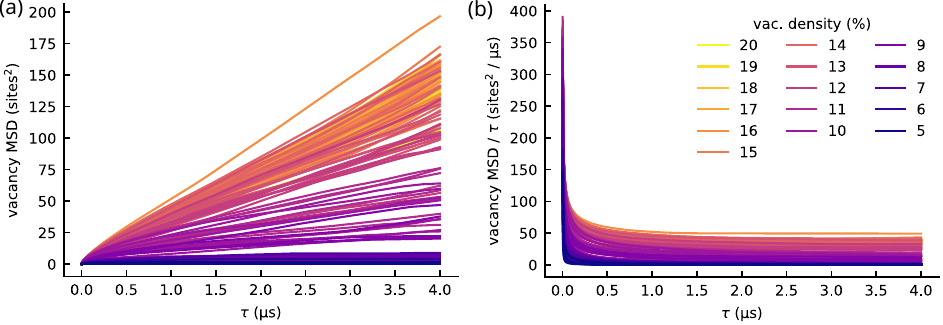}
    \caption{
        Sulfur vacancy \ac{MSD} as a function of lag time (see Eq.~4 of the main article) for different vacancy concentrations. 
For each concentration, ten \ac{KMC} simulations with a duration of 5~\textmu s were performed. 
The diffusion coefficients reported in Fig.~5 of the main manuscript were obtained from linear regressions of the \acp{MSD} (panel (a)) in the range $2.5~\text{\textmu s} < \tau < 3.5~\text{\textmu s}$, according to Eq.~5 in the main manuscript.
The corresponding $\mathrm{MSD}/\tau$ curves shown in panel (b) confirm the linear behavior in this time interval.
    }
    \label{fig:msdvacdenssweep}
\end{figure}

The diffusion coefficients reported in this work were extracted from \ac{KMC} simulations by analyzing the vacancy \ac{MSD} according to Eqs.~4 and 5 of the main manuscript. 
This approach is valid only in the long-lag-time limit, where diffusive motion is characterized by a linear increase of the \ac{MSD} with time, corresponding to a constant $\mathrm{MSD} / \tau$. 
At shorter times, deviations from this linear behavior may occur due to transient or constrained dynamics. 
As demonstrated in Fig.~\ref{fig:msdvacdenssweep}, the vacancy \acp{MSD} reach the linear, diffusive regime within the 5~\textmu s duration of the \ac{KMC} simulations reported in Figs.~4 and 5 of the main manuscript, ensuring that the calculated diffusion coefficients are well converged.

\medskip
\FloatBarrier
\section{Links to additional material}
Videos corresponding to KMC  calculations underlying Fig.~3 and Fig.~S4 can be found at: \url{https://doi.org/10.71758/refodat.88}

The KMC source code used in this study is available at: \url{https://github.com/afloetotto/mos2_montecarlo}

\bibliography{exportNoAbstract}